\documentclass[journal]{IEEEtran}

\ifCLASSINFOpdf
\else
   \usepackage[dvips]{graphicx}
\fi
\usepackage{url}
\usepackage[hidelinks]{hyperref}
\usepackage{scalerel}
\usepackage{authblk}
\usepackage{cite}
\usepackage{graphicx}
\usepackage{subcaption}
\usepackage{bm}
\usepackage{algorithm}
\usepackage{algpseudocode}
\usepackage{comment}
\usepackage{amsmath,amsfonts} 
\usepackage{amssymb}
\usepackage{mathrsfs}
\usepackage{amsthm}



\usepackage{flushend}

\usepackage{caption}
\begin{document}
\title{Local Ambiguity Shaping for Doppler-Resilient Sequences Under Spectral and PAPR Constraints}
 

\author{
    \IEEEauthorblockN{
        Shi He\IEEEauthorrefmark{1},
        Lingsheng Meng\IEEEauthorrefmark{1}\IEEEauthorrefmark{3},
        Yao Ge\IEEEauthorrefmark{3},
        Yong Liang Guan\IEEEauthorrefmark{1}\IEEEauthorrefmark{3},
        David Gonz\'alez G.\IEEEauthorrefmark{4},
        and Zilong Liu\IEEEauthorrefmark{5}
    }

    \IEEEauthorblockA{\small\IEEEauthorrefmark{1}School of Electrical and Electronics Engineering, Nanyang Technological University, Singapore, \{he0001hi, meng0071\}@e.ntu.edu.sg}

    \IEEEauthorblockA{\small\IEEEauthorrefmark{3}Continental-NTU Corporate Lab, Nanyang Technological University, Singapore, \{yao.ge, eylguan\}@ntu.edu.sg}

    \IEEEauthorblockA{\small\IEEEauthorrefmark{4}Continental Automotive Technologies GmbH, Germany, david.gonzalez.g@ieee.org}

    \IEEEauthorblockA{\small\IEEEauthorrefmark{5}School of Computer Science and Electronics Engineering, University of Essex, United Kingdom, zilong.liu@essex.ac.uk}\vspace{-1cm}
}
\renewcommand\Authsep{, }
\renewcommand\Authands{, and }
\maketitle
\vspace{-0.3cm}
\begin{abstract}
This paper focuses on designing Doppler-resilient sequences with low local Ambiguity Function (AF) sidelobes, subject to certain spectral and Peak-to-Average Power Ratio (PAPR) constraints. To achieve this, we propose two distinct optimization algorithms: (i) an Alternating Minimization (AM) algorithm for superior Weighted Peak Sidelobe Level (WPSL) minimization, and (ii) a low-complexity Augmented Lagrangian-assisted Majorization Minimization (ALaMM) algorithm with effective WPSL suppression. The proposed schemes hold great potential for sequence design in future 6G and integrated sensing and communication applications, supporting robust sensing under spectral coexistence constraints in high-mobility scenarios.
\end{abstract}
\vspace{-0.6cm}
\IEEEpeerreviewmaketitle
\vspace{-0.5cm}
\section{Introduction}
\vspace{-0.2cm}
\IEEEPARstart{T}HE rise of Integrated Sensing and Communication (ISAC) and high-mobility communications drive the demand for Doppler-resilient sequences in future 6G systems \cite{B6G_3,B6G_1}. A key tool in sequence analysis is the Ambiguity Function (AF), serving a critical role in delay-Doppler estimation for determining target distance, relative velocity, and distinguishing multiple targets \cite{deng2025,Meng2025,chirp}. 

In practical scenarios, it is desirable to shape local AF within certain delay-Doppler zone of operation, as the maximum delay and Doppler shift are typically much smaller than the signal bandwidth and sequence duration \cite{signalbw}. Additionally, to enhance power transmission efficiency, sequences with low Peak-to-Average Power Ratio (PAPR) values are preferred \cite{wu2017transmit}. Extensive research has focused on designing unimodular sequences with a thumbtack-type local AF by suppressing the Weighted Integrated Sidelobe Level (WISL) under certain PAPR constraints via, for example, Majorization-Minimization (MM) approach and its variants \cite{song2015sequence,wang2022joint,localAFshaping}. Recent studies highlight that reducing the Weighted Peak Sidelobe Level (WPSL) of AF leads to improved detection performance as it can mitigate the risk of weak targets being masked by the sidelobes of stronger ones \cite{psl,sdr2023}.

In future 6G networks, multiple Electromagnetic (EM) systems may need to coexist with limited spectral resources \cite{B6G_3}. However, the design of Doppler-resilient communication or radar sequences with low WPSL under spectral constraints is not well explored in the literature. A recent study \cite{sdr2023} proposed a Semidefinite Relaxation (SDR) based method that minimizes the Peak Sidelobe Level (PSL) of auto-correlation for unimodular sequences under spectral constraints, but the effect of Doppler shift is not considered.

Inspired by \cite{sdr2023}, this paper first proposes an Alternating Minimization (AM) algorithm based on SDR for local AF shaping under spectral constraints. Our study shows that it can achieve low WPSL but at the cost of high computational complexity. We then propose a low-complexity Augmented Lagrangian-assisted Majorization Minimization (ALaMM) algorithm, while maintaining competitive performance. Unlike prior works that impose strict unimodular constraints, the proposed methods allow for more flexible PAPR control, offering greater adaptability to various application scenarios. Numerical results verify the effectiveness of the proposed algorithms. Specifically, the optimized sequences achieve stopband attenuation comparable to that of the filtered random polyphase sequence obtained via stopband filtering \cite{firfilter}, while attaining lower sidelobe than the chirp sequence \cite{chirp}, whose PSL is known to be asymptotically order-optimal within certain zones of operation \cite{B6G_1, bound2024}.
\vspace{-0.40cm}
\section{Problem Formulation}
\vspace{-0.20cm}
In this section, we formulate the sequence design problem for minimizing WPSL in local AF while satisfying spectral and PAPR constraints. For a complex sequence of length $N$, $ \boldsymbol{x}=[x_1, \cdots, x_{N}]^T$, where $(\cdot)^T$ denotes the transpose operator, the discrete WPSL associated with $\boldsymbol{x}$ over the delay-Doppler zone of operation $\boldsymbol{\Gamma}$ can be expressed as \cite{wang2022joint},\cite{localAFshaping}: 
\begin{equation}
    \text{WPSL}=\max_{(k,f_l) \in \boldsymbol{\Gamma}}\left\{w_k\left|\boldsymbol{x}^{\dagger}\boldsymbol{U}_{k,l}\boldsymbol{x}\right|\right\},
    \label{WPSL:org}
\end{equation} where $w_k$ is the weight of the $k$th delay bin, $\boldsymbol{\Gamma}=\{\left(k,f_l\right) \mid k \in [-r,r], f_l \in [f_\text{lower}, f_\text{upper}], (k,f_l) \neq (0,0)\}$ represents the zone of operation with $r$ being the maximum normalized delay bin. $\left[f_\text{lower}, f_\text{upper}\right]$ denotes the normalized Doppler frequency interval in $\boldsymbol{\Gamma}$, and is uniformly discretized into $L$ bins, with $f_l=f_\text{lower}+(l-1)\cdot\Delta f,l=1,\cdots,L$. Note that $\Delta f=\left(f_\text{upper}-f_\text{lower}\right)/\left(L-1\right)$ can take any real number. The symbol $(\cdot)^{\dagger}$ denotes the conjugate transpose, and $|\cdot|$ is the absolute value operator. 
The matrix $\boldsymbol{U}_{k,l}=\mathbf{J}_{k}\textbf{Diag}\left(\boldsymbol{p}\left(f_l\right)\right)$ where $\mathbf{J}_{k}$ is the shift matrix defined as:
\begin{equation}
    \notag
    \mathbf{J}_k(n,m)=\Big\{\begin{array}{ll}1,&n-m=k;\\0,&n-m\neq k, \end{array}
\end{equation}
and $\boldsymbol{p}(f_l)=[e^{j2\pi f_l},\ldots,e^{j2\pi Nf_l}]^T$ represents the normalized Doppler frequency vector. $\textbf{Diag}\left(\boldsymbol{p}(f_l)\right)$ denotes a diagonal matrix with vector $\boldsymbol{p}(f_l)$ filling its principal diagonal.

We incorporate spectral constraints into the design process to ensure that the sequence maintains low energy levels in designated stopbands \cite{sdr2023}. Given the stopband frequency interval in the normalized spectrum, $\mathscr{F}_{\mathrm{stop}}\subseteq[0,1]$, we aim to control the Energy Spectral Density (ESD) of sequence $\boldsymbol{x}$ over $\mathscr{F}_{\mathrm{stop}}$. In practice, we discretize $\mathscr{F}_{\mathrm{stop}}$ into $N_{f}$ frequency points, forming the set $S=\big\{1,\cdots,N_f\big\}$. For the $s$th frequency bin in $\mathscr{F}_{\mathrm{stop}}$, we define the Discrete Time Fourier Transform (DTFT) vector $\boldsymbol{f}_s=\left[1 \ e^{j(2\pi f_s)\times1} \cdots e^{j(2\pi f_s)\times(N-1)}\right]^T$, and the matrix associated with it, $\mathbf{F}_s=\boldsymbol{f}_s\boldsymbol{f}_s^{\dagger}$. Then the energy at normalized frequency $f_s$ can be represented as
\begin{equation} \notag
\left|X\left(f_s\right)\right|^2=\left|\sum_{n=0}^{N-1}x_ne^{-j2\pi f_s n}\right|^2=\boldsymbol{x}^\dagger\mathbf{F}_s\boldsymbol{x}.    
\end{equation}
Thus, the spectral constraints can be expressed as $\boldsymbol{x}^\dagger\mathbf{F}_s\boldsymbol{x}\leq U_{\max},\forall s\in S,$
where $U_{\max}$ is the maximum allowable energy level at $ f_s $, defined as $ U_{\max}=N\times 10^{-0.1\times A}$, and $A$ is the desired stopband attenuation in decibel (dB) \cite{sdr2023}.

Besides, to restrain the PAPR value in the design process, we introduce the constraint as $\left|x_n\right|\leq \sqrt{\gamma}, \ n=1,\cdots,N,$ where $\left|x_n\right|$ is the modulus of the $n$th element of sequence $\boldsymbol{x}$ and $\gamma\in\left[1,N\right)$ is the maximum allowable PAPR \cite{wu2017transmit}.

Finally, the WPSL minimization problem under spectral and PAPR constraints is formulated as follows,
\begin{equation}
\begin{aligned}\label{WPSL_opt}
&\min_{\boldsymbol{x}}& &\hspace{-0.25em} \max_{(k,f_l) \in \boldsymbol{\Gamma}}\left\{w_k\left|\boldsymbol{x}^\dagger\boldsymbol{U}_{k,l}\boldsymbol{x}\right|\right\} \\
&\ \mathrm{s.t.}& &\boldsymbol{x}^{\dagger}\boldsymbol{x}=N,\\
& & &|x_n|\leq \sqrt{\gamma},\ n=1,\cdots,N,\\
& & &\boldsymbol{x}^\dagger\mathbf{F}_s\boldsymbol{x}\leq U_{\max},\forall s\in S.\end{aligned}\end{equation} 
The nonconvex nature of both the objective function and the energy constraint makes problem (\ref{WPSL_opt}) NP-hard \cite{localAFshaping}. In the subsequent section, we present the proposed algorithms to address this problem in detail.
\vspace{-0.4cm}
\section{Proposed Algorithms}
\vspace{-0.2cm}
In this section, we introduce the proposed AM and ALaMM algorithms, respectively.
\vspace{-0.4cm}
\subsection{Proposed AM Algorithm}
\label{section:AM algorithm}
Inspired by \cite{sdr2023}, we reformulate problem (\ref{WPSL_opt}) as its epigraph as follows:
\begin{equation}
\label{opt:epigraph}
\begin{aligned}
&\operatorname*{min}_{\boldsymbol{x}, \varphi} &\varphi\ & &\ \\
&\ \mathrm{s.t.} &\mathrm{Tr}&\left(w_k\boldsymbol{U}^{\dagger}_{k,l}\boldsymbol{x}\boldsymbol{x}^{\dagger}\boldsymbol{U}_{k,l}\boldsymbol{x}\boldsymbol{x}^{\dagger}\right)\leq \varphi, \forall \left(k,f_l\right)\in \boldsymbol{\Gamma}, &\ \\
&\ &\mathrm{Tr}&\left(\mathbf{F}_s\boldsymbol{x}\boldsymbol{x}^{\dagger}\right)\leq U_{\max}, \forall s\in S, &\ \\
&\ &\mathrm{Tr}&\left(\boldsymbol{x}\boldsymbol{x}^{\dagger}\right)=N, &\ \\
&\ &\mathrm{Tr}&\left(\boldsymbol{E}_n\boldsymbol{x}\boldsymbol{x}^{\dagger}\right)\leq \gamma, n=1,\cdots,N,&\
\end{aligned}
\end{equation} where $\varphi$ is an auxiliary variable, $\mathrm{Tr}(\cdot)$ denotes the trace of a matrix, and $\boldsymbol{E}_{n}=\boldsymbol{e}(n)\boldsymbol{e}(n)^{T}$, with $\boldsymbol{e}(n)$ being the $n$th N-dimentional standard vector. Then, problem (\ref{opt:epigraph}) is reformulated into an equivalent biconvex problem as follows,
\begin{equation}
\label{opt:equ_biconvex}
\begin{aligned}
&\operatorname*{min}_{\boldsymbol{X}_1, \boldsymbol{X}_2, \varphi} &\varphi \ & & \\
&\ \ \ \mathrm{s.t.} &\mathrm{Tr}&\left(w_k\boldsymbol{U}^{\dagger}_{k,l}\boldsymbol{X}_1\boldsymbol{U}_{k,l}\boldsymbol{X}_2\right)\leq \varphi, \forall \left(k,f_l\right)\in \boldsymbol{\Gamma}, & \\
&\ &\mathrm{Tr}&\left(\mathbf{F}_s\boldsymbol{X}_1\right)\leq U_{\max},\forall s\in S, & \\
&\ &\mathrm{Tr}&\left(\mathbf{F}_s\boldsymbol{X}_2\right)\leq U_{\max},\forall s\in S, & \\
&\ &\mathrm{Tr}&\left(\boldsymbol{X}_1\right)=N, \; \mathrm{Tr}\left(\boldsymbol{X}_2\right)=N, & \\
&\ &\mathrm{Tr}&\left(\boldsymbol{E}_n\boldsymbol{X}_1\right)\leq \gamma, n=1,\cdots,N,& \\
&\ &\mathrm{Tr}&\left(\boldsymbol{E}_n\boldsymbol{X}_2\right)\leq \gamma, n=1,\cdots,N,& \\
&\ &\mathrm{Tr}&\left(\boldsymbol{\boldsymbol{X}_1}\boldsymbol{\boldsymbol{X}_2}\right)=\mathrm{Tr}\left(\boldsymbol{\boldsymbol{X}_1}\right)\mathrm{Tr}\left(\boldsymbol{\boldsymbol{X}_2}\right),
\end{aligned}
\end{equation}
where $\boldsymbol{X}_1=\boldsymbol{X}_2=\boldsymbol{x}\boldsymbol{x}^{\dagger}$ is the autocorrelation matrix of $\boldsymbol{x}$.
By introducing the penalty function: $\mathrm{Tr}\left(\boldsymbol{\boldsymbol{X}_1}\right)\mathrm{Tr}\left(\boldsymbol{\boldsymbol{X}_2}\right)-\mathrm{Tr}\left(\boldsymbol{\boldsymbol{X}_1}\boldsymbol{\boldsymbol{X}_2}\right)=N^2-\mathrm{Tr}\left(\boldsymbol{\boldsymbol{X}_1}\boldsymbol{\boldsymbol{X}_2}\right)$, we can break problem (\ref{opt:equ_biconvex}) into the following two subproblems,
\begin{equation}
\label{opt:biconvex_1}
\begin{aligned}
&\operatorname*{min}_{\boldsymbol{X}_1, \varphi} &(1&-\eta)\varphi+\eta\left[N^2-\mathrm{Tr}\left(\boldsymbol{X}_1\boldsymbol{X}_2\right)\right]^2 &\ \\
&\ \mathrm{s.t.} &\mathrm{Tr}&\left(w_k\boldsymbol{U}^{\dagger}_{k,l}\boldsymbol{X}_1\boldsymbol{U}_{k,l}\boldsymbol{X}_2\right)\leq \varphi, \forall \left(k,f_l\right)\in \boldsymbol{\Gamma}, &\ \\
&\ &\mathrm{Tr}&\left(\mathbf{F}_s\boldsymbol{X}_1\right)\leq U_{\max},\forall s\in S, &\ \\
&\ &\mathrm{Tr}&\left(\boldsymbol{X}_1\right)=N, &\ \\
&\ &\mathrm{Tr}&\left(\boldsymbol{E}_n\boldsymbol{X}_1\right)\leq \gamma, n=1,\cdots,N,&\
\end{aligned}
\end{equation}
\begin{equation}
\label{opt:biconvex_2}
\begin{aligned}
&\operatorname*{min}_{\boldsymbol{X}_2, \varphi} &(1&-\eta)\varphi-\eta\cdot\mathrm{Tr}(\boldsymbol{X}_1\boldsymbol{X}_2) &\ \\
&\ \mathrm{s.t.} &\mathrm{Tr}&\left(w_k\boldsymbol{U}^{\dagger}_{k,l}\boldsymbol{X}_1\boldsymbol{U}_{k,l}\boldsymbol{X}_2\right)\leq \varphi, \forall \left(k,f_l\right)\in \boldsymbol{\Gamma}, &\ \\
&\ &\mathrm{Tr}&\left(\mathbf{F}_s\boldsymbol{X}_2\right)\leq U_{\max},\forall s\in S, &\ \\
&\ &\mathrm{Tr}&\left(\boldsymbol{X}_2\right)=N, &\ \\
&\ &\mathrm{Tr}&\left(\boldsymbol{E}_n\boldsymbol{X}_2\right)\leq \gamma, n=1,\cdots,N,&\
\end{aligned}
\end{equation}
where $\eta \in [0,1]$ is a preset weight. It is easy to verify subproblems (\ref{opt:biconvex_1}) and (\ref{opt:biconvex_2}) are convex, and can be solved efficiently by the CVX package. The proposed AM algorithm is summarized in \textbf{Algorithm} \ref{algm_1_AM}, where $\varepsilon_{r}$ and $\varepsilon_{x}$ check if the solution converges to a rank-one solution, and $t_{\max}$ is the maximum experiment time. The computational complexity of \textbf{Algorithm} \ref{algm_1_AM} is approximately $\mathcal{O}\left(2\times\left(2\times r\cdot L+N_f\right)\times N^6\right)$. 
\begin{algorithm}[!t]
\caption{Proposed AM Algorithm}
\label{algm_1_AM}
\begin{algorithmic}[1]
\Require $\boldsymbol{x}_\text{init}, N, w_k, \boldsymbol{\Gamma}, \eta, N_f, U_{\max}, \mathscr{F}_\mathrm{stop}, \varepsilon_x, \varepsilon_{r}, \boldsymbol{\gamma}, t_{\max};$
\Ensure Optimized sequence $\boldsymbol{x}_\text{opt}$;
\State \textbf{Initialization}: Let $\boldsymbol{X}_{2}^{(0)} = \boldsymbol{x}_\text{init} \boldsymbol{x}_\text{init}^{\dagger}$, $t = 0$;
\State Compute $\{\boldsymbol{U}_{k,l}\}_{k,l\in \boldsymbol{\Gamma}}$ and $\{\boldsymbol{E}_n\}_{n \in \{1,\dots,N\}}$;
\Repeat
    \State Solve the problem (\ref{opt:biconvex_1}) for $\boldsymbol{X}_1^{(t+1)}$ using CVX, fixing $\boldsymbol{X}_2 = \boldsymbol{X}_2^{(t)}$;
    \State Solve the problem (\ref{opt:biconvex_2}) for $\boldsymbol{X}_2^{(t+1)}$ using CVX, fixing $\boldsymbol{X}_1 = \boldsymbol{X}_1^{(t+1)}$;
    \State $t \gets t + 1$;
\Until{
    \[
    \frac{\operatorname{tr}\left(\boldsymbol{X}_1\right)\operatorname{tr}\left(\boldsymbol{X}_2\right) - \operatorname{tr}\left(\boldsymbol{X}_1\boldsymbol{X}_2\right)}
    {\operatorname{tr}\left(\boldsymbol{X}_1\right)\operatorname{tr}\left(\boldsymbol{X}_2\right)}
    = 1 - \frac{\operatorname{tr}\left(\boldsymbol{X}_1\boldsymbol{X}_2\right)}{N^2} \leq \varepsilon_x
    \]
    \textbf{or} $t \geq t_{\max}$;
}
\State Perform singular value decomposition on $\boldsymbol{X}_2^{(t)}$: $\boldsymbol{X}_2 = \boldsymbol{U} \boldsymbol{\Sigma} \mathbf{V}^\dagger$;
\State Set $\sigma_0 = \Sigma_{0,0}$ and $\sigma_1 = \Sigma_{1,1}$;
\If{$\sigma_1 / \sigma_0 \leq \varepsilon_{r}$}
    \State Obtain $\boldsymbol{x}_\text{opt} = \sqrt{\sigma_0} \cdot \boldsymbol{U} \boldsymbol{e}_0$;
\Else
    \State Declare failure of convergence.
\EndIf
\end{algorithmic}
\end{algorithm}
\vspace{-0.4cm}
\subsection{Proposed ALaMM Algorithm}
\label{section:ALaMM algorithm}
To tackle problem (\ref{WPSL_opt}) with lower complexity, we approximate the WPSL metric in (\ref{WPSL:org}) by using the \( l_p \)-norm with \( p \geq 2 \) as follows \cite{wang2022joint,song2015sequence}:\vspace{-0.2cm}
\begin{equation}
f(\boldsymbol{x})=\sum_{k\in\boldsymbol{\Gamma}}\sum_{l=1}^L w_k\left|\boldsymbol{x}^\dagger \boldsymbol{U}_{k,l}\boldsymbol{x}\right|^p,
\label{eqn:lp-obj}
\end{equation}
and with larger $p$, the approximation is more accurate. Then, problem (\ref{WPSL_opt}) is transformed to 
\begin{equation}\label{lp_norm_opt}
\begin{aligned}
&\min_{\boldsymbol{x}}&\ &f\left(\boldsymbol{x}\right) \\
&\ \mathrm{s.t.}& &\boldsymbol{x}^{\dagger}\boldsymbol{x}=N,\\
&\ &\ &\left|x_n\right|\leq\sqrt{\gamma},\ n=1,\cdots,N,\\
&\ &\ &\boldsymbol{x}^{\dagger}\mathbf{F}_s\boldsymbol{x}\leq U_\text{max},\forall s\in S.
\end{aligned}
\end{equation}
With $p\to \infty$, problem (\ref{lp_norm_opt}) is equivalent to problem (\ref{WPSL_opt}). 
We develop the augmented Lagrangian of problem (\ref{lp_norm_opt}) as follows,
\begin{align} 
\label{eqn:Augmented-Lagrangian}
\mathcal{L}_\rho({\boldsymbol{x}},\boldsymbol{\lambda})=f(\boldsymbol{x})&+\sum_{s \in S}\lambda_{s}\left(\boldsymbol{x}^\dagger\mathbf{F}_s\boldsymbol{x}-U_\text{max}\right)\notag \\
&+\frac\rho2\sum_{s \in S}\left(\boldsymbol{x}^\dagger\mathbf{F}_s\boldsymbol{x}-U_\text{max}\right)^2,
\end{align}
where $\rho>0$ and $\boldsymbol{\lambda}$ are the step size and vector of Lagrange multipliers, respectively, with $\lambda_{s}$ being the $s$th element of $\boldsymbol{\lambda}$. Thus, solving the optimization problem (\ref{lp_norm_opt}) falls into the following two steps:
1) With the obtained $ \boldsymbol{\lambda}^{(t)} $ at the $t$th iteration, update $\boldsymbol{x}^{(t+1)}$ by solving
\begin{equation}
\begin{aligned}
&\min_{\boldsymbol{x}}&\ &\mathcal{L}_{\rho}\left(\boldsymbol{x},\boldsymbol{\lambda}^{(t)}\right)\\ 
&\ \mathrm{s.t.}&&\boldsymbol{x}^{\dagger}\boldsymbol{x}=N,\\
&\ &&|x_n|\leq\sqrt{\gamma},\ n=1,\cdots,N.\end{aligned}
\label{p:Augmented-Lagrangian}\end{equation}

2) With the obtained $ \boldsymbol{x}^{(t+1)}$, vector $\boldsymbol{\lambda}^{(t+1)}$ is updated by: ${\lambda}_{s}^{(t+1)}={\lambda}_{s}^{(t)}+\rho\left(\left(\boldsymbol{x}^{(t+1)}\right)^\dagger\mathbf{F}_{s}\boldsymbol{x}^{(t+1)}-U_\text{max}\right), \forall s \in S$, 
where $ \lambda_{s}^{(t+1)} $ is the $s$th element of Lagrangian multiplier vector $\boldsymbol{\lambda}^{(t+1)}$ at the $(t+1)$th iteration.

In step 1), we ignore the irrelevant terms and rewrite problem (\ref{p:Augmented-Lagrangian}) as following, 
\begin{equation}
    \begin{aligned}
    \label{opt:simplified}
&\min_{\boldsymbol{x}}& &f\left(\boldsymbol{x}\right)+\sum_{s\in S}\left(\lambda_{s}^{(t)}-\rho U_\text{max}\right)\boldsymbol{x}^{\dagger}\mathbf{F}_s\boldsymbol{x}\\
&& &+\frac\rho2\sum_{s\in S}\left|\boldsymbol{x}^\dagger\mathbf{F}_s\boldsymbol{x}\right|^2 \\
&\ \mathrm{s.t.}& &\boldsymbol{x}^{\dagger}\boldsymbol{x}=N,\\
&\ & &|x_n|\leq\sqrt{\gamma},\ n=1,\cdots,N.
\end{aligned}
\end{equation}
The objective function of problem (\ref{opt:simplified}) can be majorized by a quadratic surrogate with constants ignored as follows \cite{wang2022joint,song2015sequence, localAFshaping},
\begin{equation} 
\label{eqn:Majorized-AL}
\begin{aligned}
&\min_{\boldsymbol{x}} & &\boldsymbol{x}^{\dagger}\boldsymbol{\Phi}\left(\boldsymbol{x}^{(t)}\right)\boldsymbol{x} \\
&\ \mathrm{s.t.}& &\boldsymbol{x}^{\dagger}\boldsymbol{x}=N, \\
&\ &&|x_n|\leq\sqrt{\gamma}, n=1,\cdots,N,
\end{aligned}
\end{equation}
with $\boldsymbol{x}^{(t)}$ fixed at the $t$th iteration, and \begin{equation} \begin{aligned} \boldsymbol{\Phi}\left(\boldsymbol{x}^{(t)}\right)&=\boldsymbol{M}\left(\boldsymbol{x}^{(t)}\right)+\boldsymbol{M}\left(\boldsymbol{x}^{(t)}\right)^{\dagger}+\boldsymbol{N}\left(\boldsymbol{x}^{(t)}\right)\\&+\boldsymbol{N}\left(\boldsymbol{x}^{(t)}\right)^{\dagger}\notag+\frac\rho2\left(\boldsymbol{P}\left(\boldsymbol{x}^{(t)}\right)+\boldsymbol{P}\left(\boldsymbol{x}^{(t)}\right)^{\dagger}\right)\notag\\&+\sum_{s \in S}\lambda_{s}^{(t)}\mathbf{F}_{s}-\rho U_\text{max}\sum_{s \in S}\mathbf{F}_s, \end{aligned}\end{equation} with 
\begin{equation}\notag
\begin{aligned}
& \boldsymbol{M}\left(\boldsymbol{x}^{(t)}\right)&&\hspace{-1em}=\ \sum_{l=1}^{L}\Bigg(\sum_{k\in\boldsymbol{\Gamma}}w_{k}a_{k,l}\left(\boldsymbol{x}^{(t)}\right)^{\dagger}\boldsymbol{U}_{k,l}^{\dagger}\boldsymbol{x}^{(t)}\boldsymbol{U}_{k,l}
\\& & &\hspace{-1em}-\  \boldsymbol{\lambda}_{\max}\left(\boldsymbol{\Lambda}_l\right)\boldsymbol{x}^{(t)}\left(\boldsymbol{x}^{(t)}\right)^{\dagger}\Bigg), \\
& \boldsymbol{N}\left(\boldsymbol{x}^{(t)}\right)& &\hspace{-1em}=\ \sum_{k \in \boldsymbol{\Gamma}}\sum_{l=1}^{L}w_{k}b_{k,l}\frac{A_{k,l}^{(t)}}{\left|A_{k,l}^{(t)}\right|}\mathbf{U}^\dagger_{k,l}, \\
& \boldsymbol{P}\left(\boldsymbol{x}^{(t)}\right)& &\hspace{-1em}=\ \sum_{s \in S}\left(\boldsymbol{x}^{(t)}\right)^{\dagger}\mathbf{F_{s}}\boldsymbol{x}^{(t)}-\lambda_\text{max}\left(\boldsymbol{L}\right)\boldsymbol{x}^{(t)}\left(\boldsymbol{x}^{(t)}\right)^{\dagger},
\end{aligned}
\end{equation} 
where $ A^{(t)}_{k,l}=\left(\boldsymbol{x}^{(t)}\right)^{\dagger} \boldsymbol{U}_{k,l}\boldsymbol{x}^{(t)} $, and
\begin{equation} \notag
    \begin{aligned}&a_{k,l}&=&\quad \frac{z^p-\left|A_{k,l}^{(t)}\right|^p-p\left|A_{k,l}^{(t)}\right|^{p-1}\left(z-\left|A_{k,l}^{(t)}\right|\right)}{\left(z-\left|A_{k,l}^{(t)}\right|\right)^2},
    \\&b_{k,l}&=&\quad p\left|A_{k,l}^{(t)}\right|^{p-1}-2a_{k,l}\left|A_{k,l}^{(t)}\right|,
    \\&z&=&\quad\Biggl(\sum_{k \in \boldsymbol{\Gamma}}\sum_{l=1}^L\left|A_{k,l}^{(t)}\right|^{p}+\sum_{s\in S}\lambda^{(t)}_{s}\left(\boldsymbol{x}^{(t)}\right)^{\dagger}\mathbf{F}_s\boldsymbol{x}^{(t)} 
    \\ & &+&\quad\sum_{s\in S}\left(\left(\boldsymbol{x}^{(t)}\right)^{\dagger}\mathbf{F}_s\boldsymbol{x}^{(t)}-U_\text{max}\right)^2\Biggr)^{\frac1p}, \\
&\boldsymbol{\Lambda}_l&=&\quad \sum_{k\in\boldsymbol{\Gamma}}w_{k}a_{k,l}\operatorname{vec}\left(\boldsymbol{U}_{k,l}\right)\operatorname{vec}\left(\boldsymbol{U}_{k,l}\right)^{\dagger}, \\
&\boldsymbol{L}&=&\quad \sum_{s\in S}\operatorname{vec}\left(\mathbf{F}_{s}\right)\operatorname{vec}\left(\mathbf{F}_{s}\right)^{\dagger},
\end{aligned}
\end{equation}
with $\operatorname{vec}(\cdot)$ representing the vectorization operator and $\lambda_\text{max}\left(\boldsymbol{A}\right)$ denoting the maximum eigenvalue of matrix $\boldsymbol{A}$. The derivation of $a_{k,l}, b_{k,l}$ and $z$ can be found in \cite{song2015sequence,localAFshaping}.
To avoid high complexity in eigenvalue decomposition, we compute $\lambda_\text{max}\left(\boldsymbol{\Lambda}_l\right)$ and $\lambda_\text{max}\left(\boldsymbol{L}\right)$ in closed-form in each iteration. According to equation (15) in \cite{song2015sequence}, $\lambda_\text{max}\left(\boldsymbol{\Lambda}_l\right)=\max_{k}\left\{w_{k}a_{k,l}\left(N-\left|k\right|\right)\right\}$. Furthermore, due to the property of $\operatorname*{vec}(\cdot)$ operator \cite{vec_op} and orthogonality of DTFT vector, we can choose $\lambda_{\max}\left(\boldsymbol{L}\right)=N^2$ since $\boldsymbol{L}\operatorname{vec}\left(\mathbf{F}_{s}\right)=\sum_{i\in S}\operatorname{vec}\left(\mathbf{F}_{i}\right)\operatorname{vec}\left(\mathbf{F}_{i}\right)^{\dagger}\operatorname{vec}\left(\mathbf{F}_{s}\right) =N^2\operatorname{vec}\left(\mathbf{F}_{s}\right)$.

Finally, problem (\ref{eqn:Majorized-AL}) can be efficiently solved resorting to the alternating projection introduced in \cite{wu2017transmit} as: $\boldsymbol{x}^{(t+1)}=\mathcal{P}\left(\boldsymbol{v}^{(t)}\right)$, where $\boldsymbol{v}^{(t)}=\left(\mu_{t}\boldsymbol{I}_{N}- \boldsymbol{\Phi}\left(\boldsymbol{x}^{(t)}\right)\right)\boldsymbol{x}^{(t)}$, with
\begin{equation}
\begin{aligned}\mu_{t}&=2N\sum_{k\in \boldsymbol{\Gamma}}\sum_{l=1}^{L}w_{k}\left(\left|a_{k,l}\left(\boldsymbol{x}^{(t)}\right)^{\dagger}\boldsymbol{U}_{k,l}^{\dagger}\boldsymbol{x}^{(t)}\right|+\left|b_{k,l}\frac{A_{k,l}^{(t)}}{\left|A_{k,l}^{(t)}\right|}\right|\right)
\notag \\ 
&+\rho N\sum_{s\in S}\left(\boldsymbol{x}^{(t)}\right)^{\dagger}\mathbf{F}_{s}^{\dagger}\boldsymbol{x}^{(t)}+\sum_{s\in S}\lambda_{s}^{(t)}N,
\label{eqn:mu}
\end{aligned}
\end{equation} and $\boldsymbol{I}_N$ represents an $N\times N$ identity matrix. Denote $m$ as the number of nonzero elements in vector $\boldsymbol{v}^{(t)}$, then $\mathcal{P}\left(\boldsymbol{v}^{(t)}\right)$ is defined as:
\begin{equation}
\label{pxt}
\begin{aligned}&\mathcal{P}\left(\boldsymbol{v}^{(t)}\right)=\boldsymbol{\chi}_{\mathbb{R}_{\geq0}}\left(N-m\gamma\right)\sqrt{\gamma}\mathbf{u}_{m}\odot e^{j\arg\left(\boldsymbol{v}^{(t)}\right)}\\&+\boldsymbol{\chi}_{\mathbb{R}_{<0}}\left(N-m\gamma\right)\min\left\{\beta|\mathbf{v}|,\sqrt{\gamma}\mathbf{1}\right\}\odot e^{j\arg\left(\boldsymbol{v}^{(t)}\right)},\end{aligned}
\end{equation}
where $\mathbf{1}$ denotes an all-ones vector, $\text{min}\{\cdot,\cdot\}$ takes the element-wise minimum of the input vectors and $\odot$ represents the Hadamard product. $|\cdot|$ and $\arg(\cdot)$ take the element-wise modulus and phase of a complex vector respectively, and $\mathbb{R}_{\geq0}$ and $\mathbb{R}_{<0}$ denote the set of non-negative and negative real numbers. In (\ref{pxt}),
\begin{equation} \notag
    \boldsymbol{\chi}_B(x)=\left\{\begin{array}{ll}1,&\mathrm{if}\ x\in B;\\0,&\text{otherwise,}\end{array}\right.
\end{equation}
\begin{equation}\mathbf{u}_m=\left[\underbrace{1,\ldots1}_m,\underbrace{\sqrt{\frac{N-m\gamma}{\left(N-m\right)\gamma}},\ldots,\sqrt{\frac{N-m\gamma}{\left(N-m\right)\gamma}}}_{N-m}\right]^{T}, \notag
\end{equation}
and
\begin{equation}
    \label{beta}
    \begin{aligned}\beta\in\Biggl\{\beta \Bigg|& \sum_{n=1}^N\min\Big\{\beta^2\left| v^{(t)}_{n}\right|^2,\gamma\Big\}=N,\\&\beta\in\left[0,\sqrt{\gamma}\Big/{\min\left\{\left|v^{(t)}_n\right|, \left|v^{(t)}_n\right|\neq0\right\}}\right]\Biggr\}.\end{aligned}
\end{equation}
$\beta$ in (\ref{beta}) can be solved by some root-finding toolbox in MATLAB easily.

The ALaMM algorithm is summarized in \textbf{Algorithm} \ref{alg:ALaMM}. In steps 2-14, we employ the SQUAREM acceleration scheme \cite{squarem}, and the function $W(\boldsymbol{x})$ in step 9 is defined as:
\begin{align}
    \notag W(\boldsymbol{x})&=\sum_{k\in\boldsymbol{\Gamma}}\sum_{l=1}^L w_k\left|\boldsymbol{x}^\dagger \boldsymbol{U}_{k,l}\boldsymbol{x}\right|^{p}+\sum_{s\in S}\boldsymbol{x}^\dagger\lambda_{s}^{(t)}\mathbf{F}_s\boldsymbol{x} \\&+\frac\rho2\sum_{s\in S}\left|\boldsymbol{x}^\dagger\mathbf{F}_s\boldsymbol{x}\right|^2-\rho U_\text{max}\sum_{s\in S}\boldsymbol{x}^\dagger\mathbf{F}_s\boldsymbol{x}.
    \notag
\end{align}
The computational complexity of the proposed ALaMM algorithm is approximately $\mathcal{O}\left(\left(2\times r\cdot L+N_f\right)\times N^2\right)$, which is much lower than that of the proposed AM algorithm.
\vspace{-0.40cm}
\begin{figure}[htbp]\centering
    \includegraphics[width=0.4\textwidth]{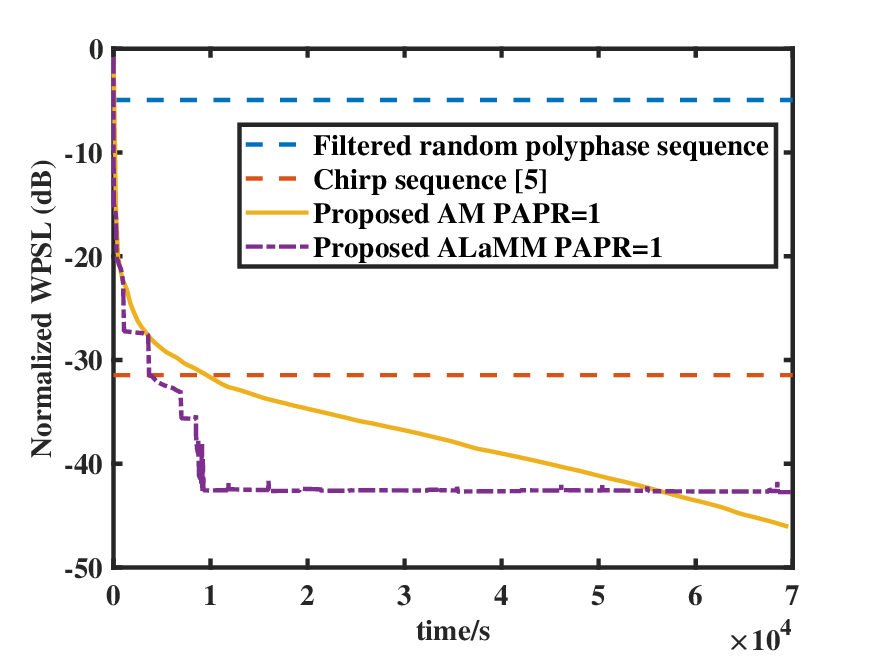}
    \caption{WPSL of different schemes versus computational time.}
    \label{fig:wpsl}
\end{figure}
\vspace{-0.35cm}
\begin{algorithm}[t]
\caption{Proposed ALaMM Algorithm}
\label{alg:ALaMM}
\begin{algorithmic}[1]
\Require $\boldsymbol{x}_\text{init}$, $\boldsymbol{\Gamma}$, $\mathscr{F}_\mathrm{stop}$, $N_f$, $\gamma$, $U_\text{max}$;
\Ensure The desired sequence $\boldsymbol{x}_\text{opt}$;
\State Let $\boldsymbol{x}^{(t)}=\boldsymbol{x}_\text{init}$, compute $\boldsymbol{v}^{(t)}$;
\Repeat
    \State Let $\boldsymbol{x}_{1} = \mathcal{P}\left(\boldsymbol{v}^{(t)}\right)$, compute $\boldsymbol{v}_1$;
    \State Let $\boldsymbol{x}_{2} = \mathcal{P}\left(\boldsymbol{v}_{1}\right)$;
    \State Compute: $\boldsymbol{r} = \boldsymbol{x}_{1} - \boldsymbol{x}^{(t)}$, $\boldsymbol{u} = \boldsymbol{x}_{2} - \boldsymbol{x}_{1} - \boldsymbol{r}$;
    \State Compute step-length $\alpha = -\left\|\boldsymbol{r}\right\|/\left\|\boldsymbol{u}\right\|$;
    \State Set $\boldsymbol{x}_{3} = \boldsymbol{x}^{(t)} - 2\alpha\boldsymbol{r} + \alpha^2\boldsymbol{u}$, compute $\boldsymbol{v}_3$;
    \State Update sequence: $\boldsymbol{x}^{(t+1)} = \mathcal{P}\left(\boldsymbol{v}_{3}\right)$;
    \While{$W\left(\boldsymbol{x}^{(t+1)}\right) > W\left(\boldsymbol{x}^{(t)}\right)$}
        \State $\alpha \gets \left(\alpha-1\right)/2$;
        \State $\boldsymbol{x}_{3} = \boldsymbol{x}^{(t)} - 2\alpha\boldsymbol{r} + \alpha^2\boldsymbol{u}$;
        \State Compute $\boldsymbol{v}_3$;
        \State Update: $\boldsymbol{x}^{(t+1)} = \mathcal{P}\left(\boldsymbol{v}_{3}\right)$;
    \EndWhile
    \State $t \gets t+1$;
\Until{exit condition is satisfied};
\State \textbf{Output:} $\boldsymbol{x}_\text{opt} = \boldsymbol{x}^{(t+1)}$.
\end{algorithmic}
\end{algorithm}
\vspace{-0.35cm}
\begin{figure}[htbp]\centering
    \includegraphics[width=0.41\textwidth]{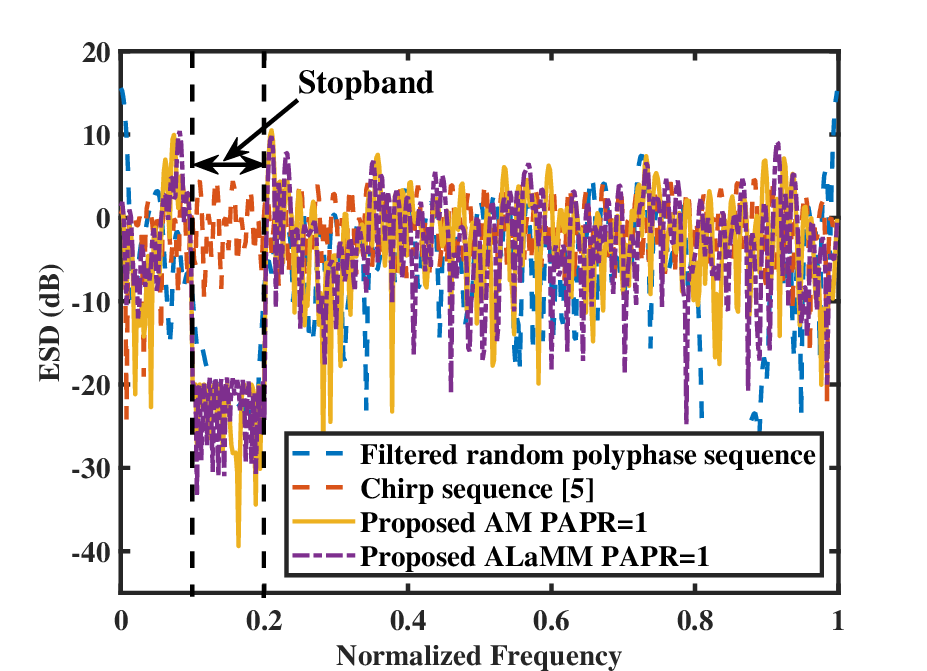}
    \caption{ESDs of different schemes.}
    \vspace{-0.2cm}
    \label{fig:ESD}
\end{figure}
\begin{figure*}[h]
    \centering
    \begin{minipage}{0.3\linewidth}
        \centering
        \includegraphics[width=\linewidth]{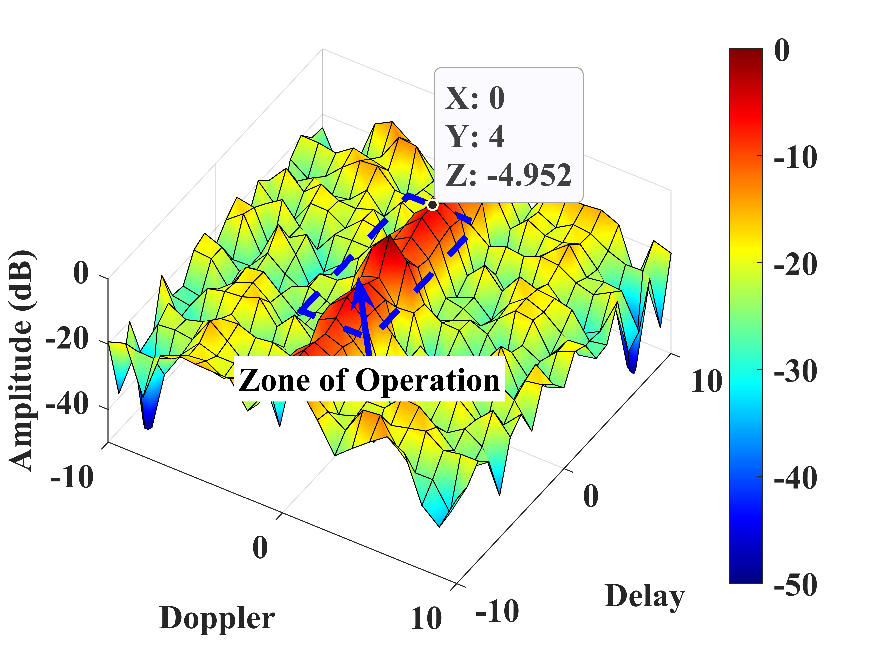}
        \subcaption{}
    \end{minipage}
    \begin{minipage}{0.3\linewidth}
        \centering
        \includegraphics[width=\linewidth]{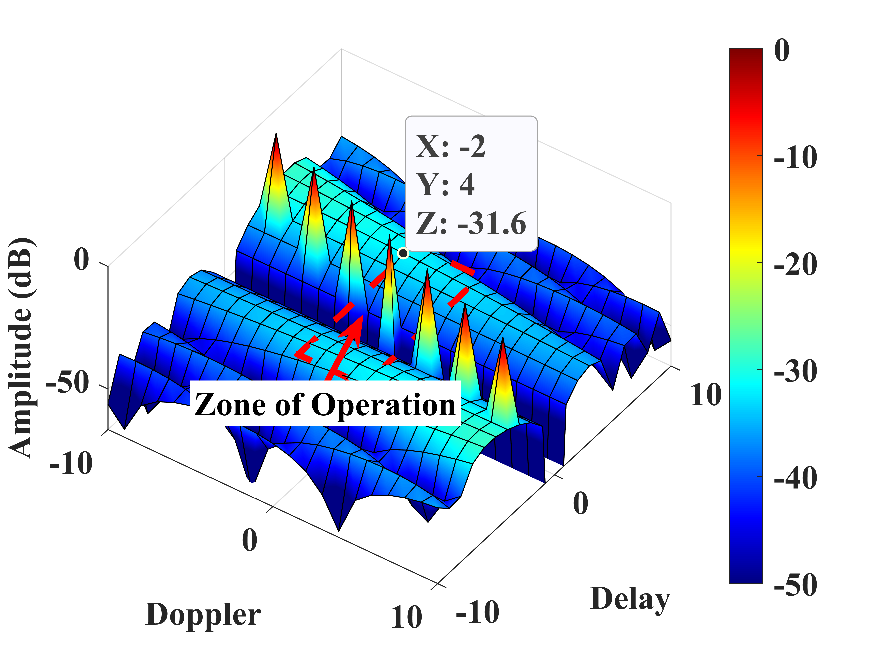}
        \subcaption{}
    \end{minipage}
    \begin{minipage}{0.3\linewidth}
        \centering
        \includegraphics[width=\linewidth]{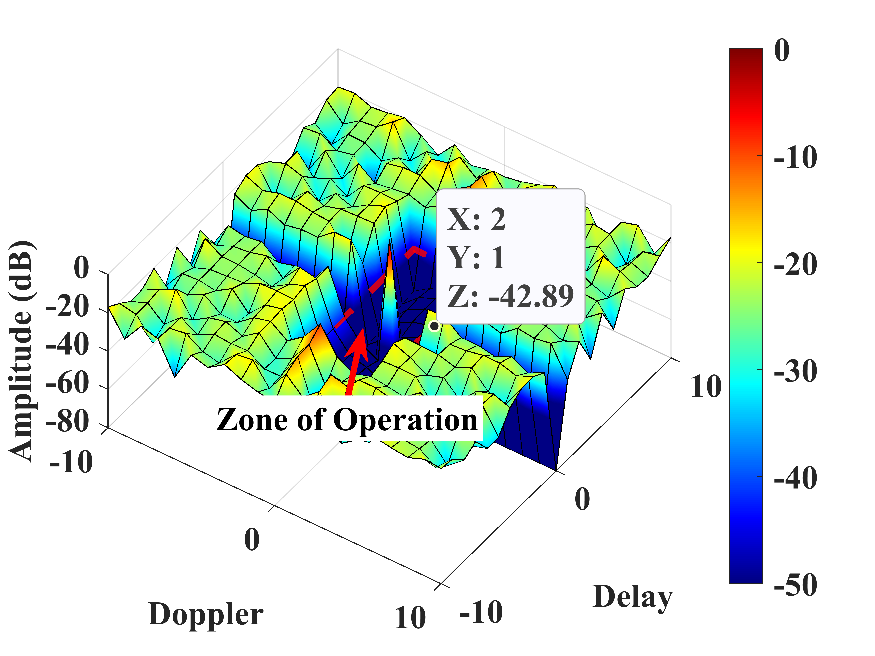}
        \subcaption{}
    \end{minipage}
    
    \begin{minipage}{0.3\linewidth}
        \centering
        \includegraphics[width=\linewidth]{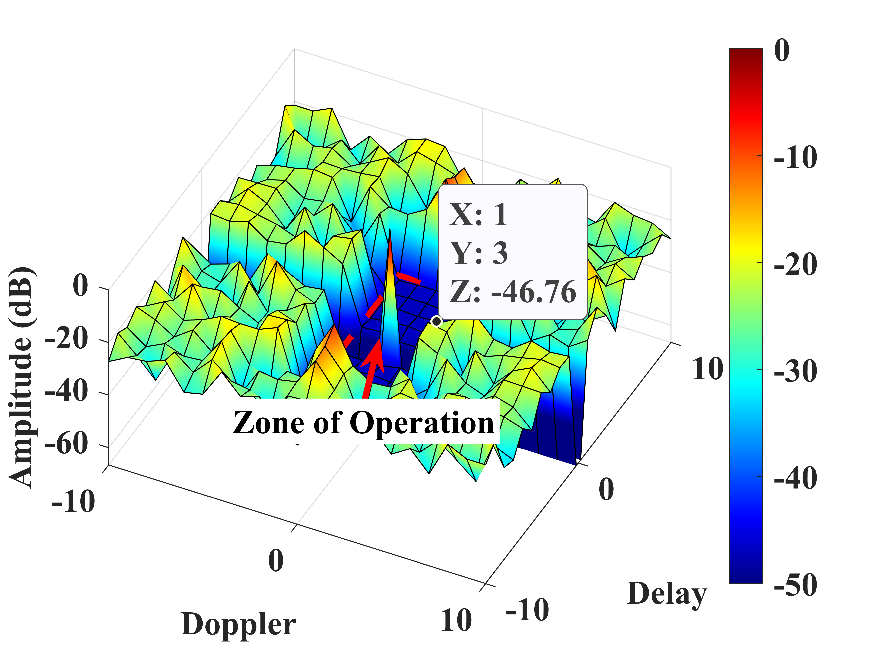}
        \subcaption{}
    \end{minipage}
    \begin{minipage}{0.3\linewidth}
        \centering
        \includegraphics[width=\linewidth]{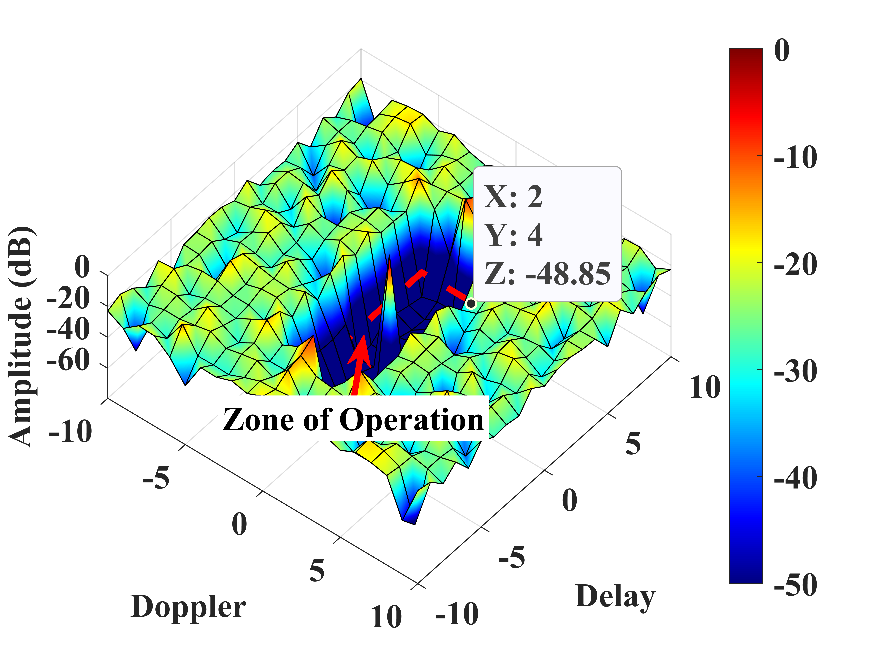}
        \subcaption{}
    \end{minipage}
    \begin{minipage}{0.3\linewidth} 
        \centering
        \includegraphics[width=\linewidth]{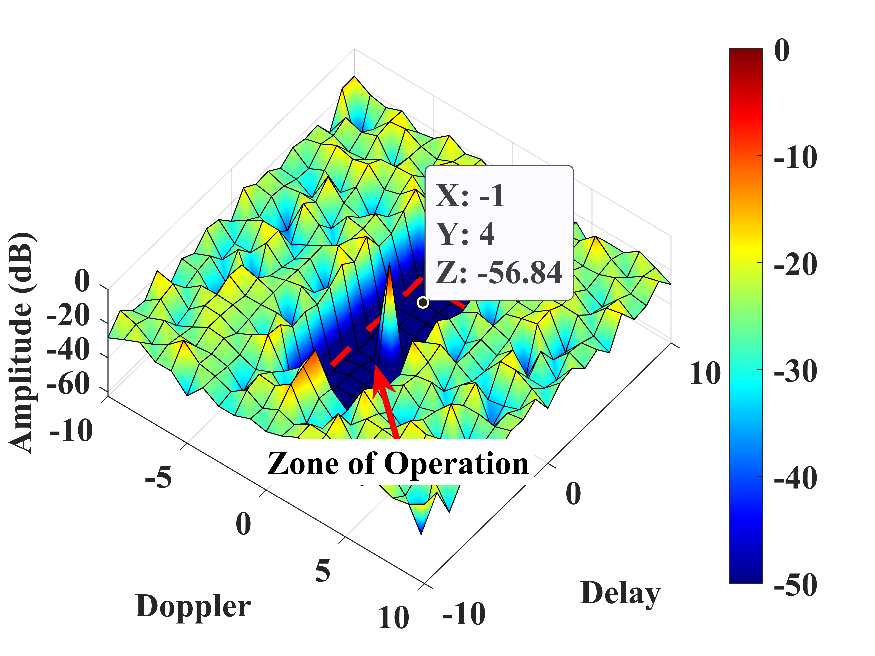}
        \subcaption{}
    \end{minipage}
    \caption{3D plot of AF over the delay-Doppler plane for (a) filtered random polyphase sequence (b) chirp sequence \cite{chirp} (c) proposed ALaMM (PAPR=1) (d) proposed AM (PAPR=1) (e) proposed ALaMM (PAPR=3) and (f) proposed AM (PAPR=3) (The optimized sequences can be found in \cite{sequence_lib}).}
    \vspace{-0.2cm}
    \label{fig:AFs}
\end{figure*}
\vspace{-0.15cm}
\section{Numerical Results}
In this section, we evaluate the performance of the proposed algorithms through numerical experiments. The generated sequences are compared with the filtered random polyphase sequence and chirp sequence to assess WPSL reduction and spectral constraint enforcement. The length of sequences here is $N=128$, and the zone of operation is $\boldsymbol{\Gamma}=\left\{(k,f_l) \mid k\in [-5,5], f_l\in[-2,2], (k,f_l)\neq (0,0)\right\}$. The normalized stopband interval is $\mathscr{F}_{\mathrm{stop}}=[0.1,0.2]$ with desired attenuation $A=20$ dB and $N_f = 50$ bins in it. The PAPR values are set as 1 and 3 respectively, and $p=22$ is chosen for the ALaMM algorithm for balanced performance and convergence time.

In Fig.~\ref{fig:wpsl}, it is observed that the unimodular sequences generated by the proposed algorithms achieve over 10 dB lower WPSL than the chirp sequence. It also implies that the proposed AM algorithm has the potential to achieve a lower WPSL than the proposed ALaMM algorithm, but at the cost of longer computational time. The ESDs shown in Fig.~\ref{fig:ESD} demonstrate that both the proposed algorithms can achieve desirable attenuation in the stopband effectively. From Fig.~\ref{fig:wpsl} and \ref{fig:ESD}, we can also conclude that, compared to the traditional filtering approach, the proposed algorithms achieve significantly lower WPSL while maintaining comparable stopband attenuation performance.

Fig.~\ref{fig:AFs} presents the 3D plots over the discrete AF plane for different schemes. Fig.~\ref{fig:AFs} (a)-(d) indicate that the proposed algorithms have a much lower WPSL than the traditional filtering approach and the chirp sequence. Additionally, Fig.~\ref{fig:AFs} (e) and (f) show that the PAPR constraint of the proposed algorithms is adjustable, and with less stringent PAPR restriction, the WPSL could be further suppressed.
\vspace{-0.3cm}
\section{Conclusion}
\vspace{-0.15cm}
In this paper, we proposed two effective algorithms for designing Doppler-resilient sequences with low WPSL while satisfying PAPR and spectral constraints. Numerical experiments show that the proposed AM algorithm achieves superior WPSL minimization and stopband attenuation but at the cost of higher computational complexity, whereas the proposed ALaMM algorithm offers greater computational efficiency.
\bibliographystyle{IEEEtran}
\bibliography{IEEEabrv,references}

\end{document}